\documentstyle[psfig]{aa}
\begin{document}
\def\fu{$f_1$}
\def\t{$\pm$}
\def\fd{$f_2$}
\def\fdu{$\phi_{21}$}
\def\fp{$f_1 + f_2$}
\def\fm{$f_2 - f_1$}
\def\cd{cd$^{-1}$}
\def\cds{cd$^{-1}$\,}
\def\kms{km~s$^{-1}$}
\def\kmss{km~s$^{-1}$\,}
\def\I{\'\i}
\def\salp{\vskip 0.3truecm}
\thesaurus{6(03.13.2, 08.15.1, 08.22.2, 10.07.3)}
\title{The light curves of the short--period variable stars 
in $\omega$ Centauri}
\author{E. Poretti}
\institute {Osservatorio Astronomico di Brera, Via Bianchi 46,
I-23807 Merate, Italy\\E--mail: poretti@merate.mi.astro.it}
\offprints{E. Poretti}
\date{Received date; Accepted Date}
\maketitle
\markboth{E. Poretti: Short--period stars in $\omega$ Cen}{ }
\begin{abstract} The Fourier decomposition was applied to the light curves of
the short period variable stars discovered by the OGLE team (Kaluzny et al. 
1996, 1997) in $\omega$ Cen. The progression of the \fdu parameter as a function
of the period is extended toward very short periods as the new values
connected directly to those of stars located in the Galaxy. However,
two groups of stars deviate: the first is located around 0.038 d and it shows
rather high \fdu values; the
second is the origin of a small change in the slope around 0.050 d. The reality
of the two features is discussed. The peculiarity of the light curve of 
OGLEGC~26 is also emphasized.
\keywords{Methods: data analysis -  Stars: oscillations - $\delta$ Sct - 
Globular clusters: $\omega$ Cen}
\end{abstract}
\section{Introduction}
Pulsating variables are being continuously discovered in the course of
large--scale projects. The Fourier decomposition describes their light
curves in a powerful, synthetic  way, supplying information on the
pulsational content.  As an example, Fourier parameters give the 
possibility to determine if a Cepheid pulsates in the fundamental or in
an overtone mode (see Pardo \& Poretti 1997 for an application to
double--mode Cepheids) and this could make  any Period--Luminosity 
relationship more clear.

The analysis of the light curve of short--period pulsating variables 
($P<$0.20 d) was carried out firstly by Antonello et al. (1986); then
Poretti et al. (1990) and Musazzi et al. (1998)
supplied new observational evidence. All these stars are located in the
Galaxy and they do not belong to clusters;  we shall call them
hereinafter ``galactic" variables. They are both Pop.~I  ($\delta$ Sct stars)
and Pop.~II (SX Phe stars) objects; no clear separation of the light curves as
a function of the population was detected.

The OGLE project collected a large amount of photometric data while 
monitoring the
globular cluster NGC~5139$\equiv\omega$ Cen (Kaluzny et al. 1996, 1997).
34 new SX Phe stars were discovered: 24 are presented  by
Kaluzny et al. (1996), 10 by Kaluzny et al. (1997).
These data can supply original results since galactic stars do not
display periods shorter than 0.06 d, while in the $\omega$ Cen sample
this value is rather an upper limit. Therefore, we have an opportunity to
verify if there is a straight connection between the two different samples
and, if any, to extend the period baseline.
\section{Period verification and refinement}
The time baseline covered by the OGLE monitoring is around 120 days
(i.e. a single observing season) for most
of the stars, but in 9 cases the available data extend over two seasons.
In this case, an improvement of the goodness of the fit could be obtained by
calculating a solution for each season and then aligning the mean magnitudes
(this procedure was applied to the measurements of OGLEGC~3, 4, 5).
As a matter
of fact, shifts up to 0.048 mag were observed in Field 5139BC, which are
surely due to observational or instrumental problems. In two cases
(OGLEGC~42, 45), we did not consider the data obtained in
one season, as they were a small part of the total and probably affected by
a misalignment which was difficult to quantify. In the remaining four cases
(OGLEGC~9, 29, 38, 59)  the
procedure of the re-alignment did not introduce appreciable effects on the fit.

We made an independent period search.
Since the baseline and the number of measurements were appropriate, all the
values previously known were confirmed. Only the case of OGLEGC~34
deserves some comment. Kaluzny et al. (1996) suspected a double--mode nature
on the basis of the period search carried out with the CLEAN algorithm.
We performed the frequency search by using the least--squares
iterative method (Vani\^cek 1971) and we obtained the power spectra shown 
in Fig.~1 (upper panel). The peak at $f$=26.1611 \cd is the highest,
but the difference
with respect to the alias at 25.1611 \cd is very small. When introducing $f$ 
as a known constituent, the power spectrum did not reveal any significant
feature in the range that would be expected for a second
period (lower panel). The CLEAN algorithm is probably responsible
for the result quoted by Kaluzny et al. (1996): because it
cannot match the odd noise distribution, the
signal is spread at different peaks.
OGLEGC~34 is probably monoperiodic, but the period is uncertain and may
be either one of the two values reported above; we have a slight preference
for $f$=26.1611 \cd because it gives
a better fit and a better residual power spectrum. Note also in the lower panel
of Fig.~1 the increasing noise at very low frequencies, the fingerprint of
night--to--night misalignments.
\begin{figure}
\centerline{\psfig{file=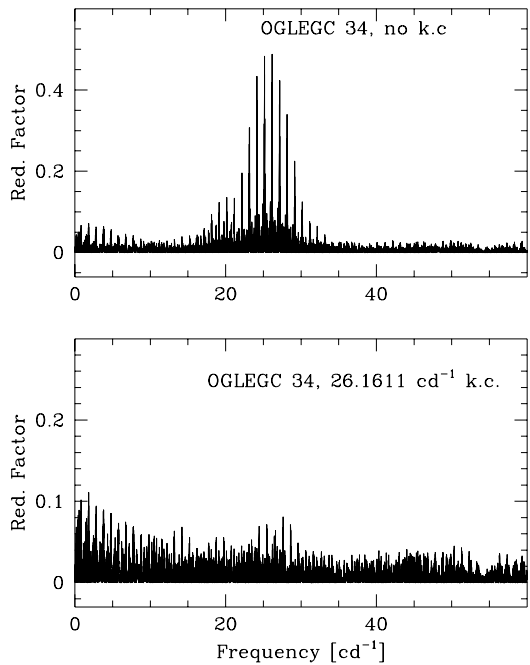,height=9truecm}}
\caption[]{The power spectra of the OGLEGC~34 measurements.
In the power spectrum shown in the upper panel the frequency $f$=26.1611 \cd
is detected as the highest peak, even if the 25.1611 \cd term is
very similar in height. After introducing the former as a known constituent
(k.c.), no significant second term is detected (lower panel). The star
has probably a single period, whose value is still ambiguous (25.1611 \cd or
26.1611 \cd)}
\end{figure}
\begin{figure}
\centerline{\psfig{file=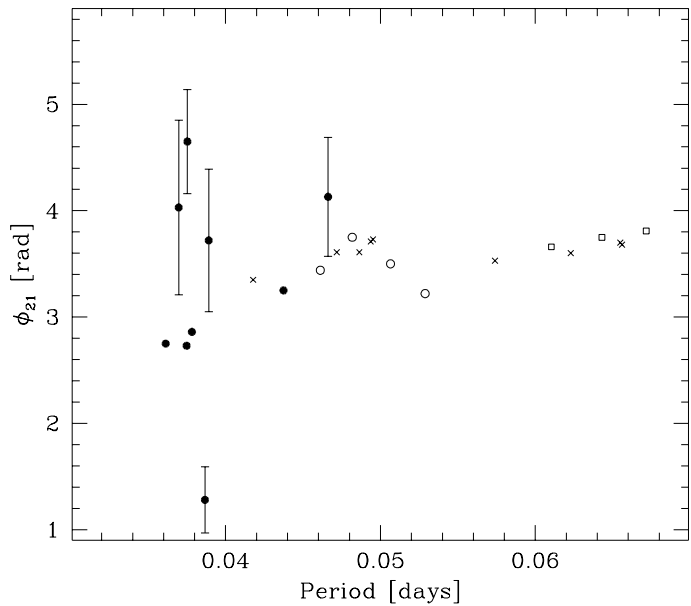,height=8.5truecm}}
\caption[]{The \fdu$-P$ progression for short period stars in $\omega$ Cen.
Different symbols for different amplitudes: filled dots for $A\leq 0.04$ mag,
open dots for 0.06$\leq$$A$$\leq$0.09 mag, crosses for $A\geq$ 0.11 mag. The
three open squares on the right side denote the three galactic stars CY Aqr, 
ZZ Mic and V831 Tau. Error bars are
reported for the individual cases discussed in the text}
\end{figure}
\begin{figure}
\centerline{\psfig{file=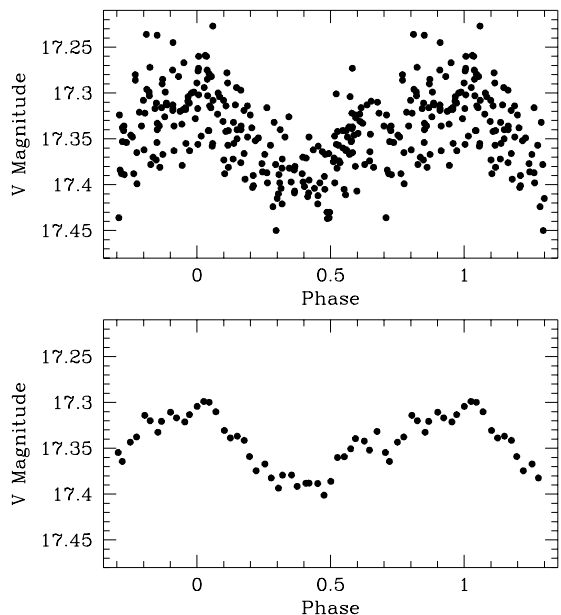,height=9truecm}}
\caption[]{The star OGLEC~26 shows an asymmetrical light curve, with
a descending branch steeper than the ascending one (individual points, upper
panel). This asymmetry is even more obvious in the mean light curve
(lower panel)}
\end{figure}
\begin{table*}
\begin{flushleft}
\caption{The phases differences ($\phi_{21}, \phi_{31}, \phi_{41}$)
 and amplitude
ratios ($R_{21}, R_{31}, R_{41}$) obtained from the Fourier decomposition
are reported together with period values. The half--amplitude of the light
variation is also reported. In 12 cases the $2f$ term (i.e. the first 
harmonic) was not detected.}
\begin{tabular}{rl rr rrrr rr}
\hline
\noalign{\smallskip}
 \multicolumn{1}{c}{OGLEGC} & \multicolumn{1}{c}{Period}& 
 \multicolumn{1}{c}{$<V>$}&  \multicolumn{1}{c}{$A_1$}&
 \multicolumn{1}{c}{$\phi_{21}$}&\multicolumn{1}{c}{$R_{21}$} &
 \multicolumn{1}{c}{$\phi_{31}$}&\multicolumn{1}{c}{$R_{31}$} &
 \multicolumn{1}{c}{$\phi_{41}$}&\multicolumn{1}{c}{$R_{41}$} \\
 \multicolumn{1}{c}{} & \multicolumn{1}{c}{[d]}& 
 \multicolumn{1}{c}{}&  \multicolumn{1}{c}{[mag]}&
 \multicolumn{1}{c}{[rad]}&\multicolumn{1}{c}{} &
 \multicolumn{1}{c}{[rad]}&\multicolumn{1}{c}{} &
 \multicolumn{1}{c}{[rad]}&\multicolumn{1}{c}{} \\
\noalign{\smallskip}
\hline
\noalign{\smallskip}
28 & 0.036134870 & 16.739 & 0.03 & 2.75 & 0.09 \\
39 & 0.036973661 & 17.558 & 0.02 & 4.03 & 0.12 \\
38 & 0.037484631 & 17.524 & 0.03 & 2.73 & 0.17 \\
36 & 0.037533733 & 17.400 & 0.04 & 4.65 & 0.09 \\
33 & 0.037826061 & 17.437 & 0.04 & 2.86 & 0.10 \\
26 & 0.038668115 & 17.348 & 0.04 & 1.28 & 0.33 \\
29 & 0.038911717 & 17.310 & 0.02 & 3.72 & 0.12 \\
25 & 0.043739175 & 17.143 & 0.03 & 3.25 & 0.13 \\
62 & 0.046620047 & 17.462 & 0.03 & 4.13 & 0.17 \\
\noalign{\smallskip}
 1 & 0.046120912 & 17.026 & 0.06 & 3.44 & 0.21 \\
 2 & 0.048181161 & 17.404 & 0.06 & 3.75 & 0.48 \\
 6 & 0.050652146 & 17.202 & 0.09 & 3.50 & 0.21 & 0.84 & 0.05 \\
27 & 0.052887667 & 17.054 & 0.06 & 3.22 & 0.10 \\
\noalign{\smallskip}
 8 & 0.041784708 & 16.749 & 0.16 & 3.35 & 0.16 \\
50 & 0.047180040 & 17.050 & 0.23 & 3.61 & 0.29 & 1.43 & 0.07 \\
32 & 0.048640261 & 16.995 & 0.13 & 3.61 & 0.27 \\
 9 & 0.049374914 & 16.943 & 0.19 & 3.71 & 0.25 & 1.28 & 0.06 & 5.30 & 0.03 \\
 4 & 0.049520847 & 16.720 & 0.15 & 3.73 & 0.27 \\
42 & 0.057399999 & 17.006 & 0.25 & 3.53 & 0.25 \\
 3 & 0.062286809 & 16.641 & 0.27 & 3.60 & 0.41 & 1.25 & 0.20 & 5.06 & 0.10 \\
 5 & 0.065491187 & 16.805 & 0.16 & 3.70 & 0.39 & 1.24 & 0.15 \\
45 & 0.065600010 & 16.849 & 0.11 & 3.68 & 0.34 & 1.15 & 0.12 \\
\noalign{\smallskip}
\hline
\end{tabular}
\end{flushleft}
\end{table*} 
\section{Fourier parameters}
As a further step, we fitted the $V$ magnitudes  by means of the formula
\begin{equation}
V(t)= V_o + \sum_i {A_i \cos [2\pi~i~f  (t-T_o) +\phi_i ]}
\end{equation}
where $f$ is the frequency, measured in cycles per day (\cd).
 From the least--squares coefficients we calculated
the Fourier parameters $R_{ij}=A_i/A_j$ (in particular $R_{21}=A_2/A_1$)
and $\phi_{ij}=j~\phi_{i}-i~\phi_{j}$ (in particular $\phi_{21}=\phi_2 -
2~\phi_1$). These parameters are reported in Tab.~1; the mean magnitude
of OGLEGC~29 is assumed from Kaluzny et al. (1996) as the values listed
in the electronic table are shifted up by 2.5 mag. The period values obtained
from the least--squares routine are listed, but they do not differ greatly
from those reported by Kaluzny et al. (1996, 1997).

Typical error bars are $\pm$0.33 rad for the \fdu values and $\pm$0.05 for the
$R_{21}$ ones.
 Note that the amplitudes quoted hereinafter are those of the
cosine terms, i.e. the half--amplitude of the light variation.
No significant $2f$ term could be evidenced in 12 cases (OGLEGC~7, 24, 34,
35, 37, 40, 46, 59, 60, 63, 66, 70). For these stars the light curves do not
deviate
appreciably from a sinusoid: that means that if a 2$^{\rm nd}$--order fit is
forced on the data, the error bar on the amplitude of the 2$f$ term is larger
than the amplitude itself. Following the same criterium, 
in 15 other cases the fit was stopped
at the 2$^{\rm nd}$--order, in 6 cases at the 3$^{\rm rd}$ and in two cases at
the 4$^{\rm th}$. 

Figure~2 shows the $\phi_{21}-P$ plot: the stars have been subdivided into three
groups according to their amplitude and different symbols have been used. As
can be noticed, there is a well defined trend in the diagrams. Moreover,
the $\phi_{21}$ values (open squares in Fig.~2)
 related to the galactic stars CY Aqr, ZZ Mic 
(Antonello et al. 1986) and V831 Tau (Musazzi et al. 1998) are in excellent
agreement with those  related to stars in $\omega$ Cen. 

There are some interesting cases:

{\it OGLEGC~26} -- The light curve is noisy (rms residual 0.033 mag), but its
shape looks quite strange, with a descending branch steeper than the
ascending one (Fig.~3, upper panel). The reality of the asymmetry is even
more obvious when considering the mean light curve (Fig.~3, lower panel).
In the Galaxy, there are two high--amplitude $\delta$ Sct stars with a similar
light curve: V1719 Cyg (Poretti \& Antonello 1988) and V798 Cyg (Musazzi et
al. 1998). Both these stars have a double--mode nature.
Since the number of measurements of OGLEGC~26 is adequate (231 on 50 nights),
a second period should be revealed by the frequency analysis, but we failed
to find it. 

{\it OGLEGC~29, 36, 39 and 62} -- There are a few cases where the \fdu
values seem to deviate from the progression described by the others (Fig.~2).
When considering the error bars, the \fdu values of the light curves
of OGLEGC~29 and 39 (3.72$\pm$0.66 rad and 4.02$\pm$0.82 rad, respectively)
are only marginally deviating; in the case of OGLEGC~62 (\fdu=4.13$\pm$0.56 
rad) the line is just within the error bar of the related point.
This discrepancy can be explained by observational scatter, since 
the amplitude of the $A_2$ terms is very small. Moreover, the error bars
may be optimistic since they are obtained from the formal error propagation.
However, we  note that the highest value (4.65$\pm$0.49 rad for OGLEGC~36)
is the more reliable one and is farther than 3$\sigma$ from the others.

\section{Discussion}
The analysis of the 34 short--period variable  stars in $\omega$ Cen 
stressed the importance of studying the Fourier parameters. The sample
of high--amplitude $\delta$ Sct and SX Phe stars is considerably enlarged
by these new variables especially toward shortest periods.
In general, many variable stars show a very small amplitude, below 0.10 mag.
Such a small value is probably responsible for the high number of sinusoidal
light curves: since the $R_{21}$ ratios are usually around 0.1, the amplitude of
the $2f$ term is very small and observational errors can mask the asymmetry
of the light curve. 

In spite of that, the \fdu parameters are confined in a narrow strip for
periods between 0.042 d and 0.07 d. Toward longer periods, there is an
overlapping with the values obtained in the case of galactic stars.
Toward shorter periods, the tendency to decreasing \fdu values  
is also verified.
It should be noted that there is a  strong difference with the results obtained by analyzing
the stars in the Carina dwarf Spheroidal Galaxy (Poretti 1999), where the
distribution is not as clear as it is here.

As a general consideration, the progression of
the \fdu parameter as  a function of the period appears in a clear
way. However, a careful analysis should take more details in
consideration:
\begin{enumerate}
\item Attention should be paid to the scatter in the distribution of the \fdu
parameters around 0.050 d in Fig.~2; in that region the mean error 
is $\pm$0.20 rad. Hence,  this intriguing feature is on the borderline to
be considered as a real change in the progression. By analogy to Cepheid
light curves (Pardo \& Poretti 1997), such a 
change can be the signature of a resonance between the fundamental mode
and  a higher overtone.

\item The small bunch of points above the progression
at 0.038 d suggests a different light curve family. Since this group of stars
shows a very small amplitude, it is possible that they are nonradial pulsators,
not necessarily radial pulsators in a higher overtone.

\item The very low \fdu value (1.28$\pm$0.31 rad) emphasizes the anomalous 
light curve of OGLEGC~26. The
fact that such a light curve is observed in a Pop.~II object is quite
surprising, since  V1719 Cyg and V798 Cyg (whose light curves are
similar) are very probably 
Pop.~I stars having a quite normal metallic content. However, their \fdu values
are higher (2.52$\pm$0.05 and 2.64$\pm$0.06 rad, respectively) and hence the
light curves are a little different. In many cases, it seems that the
phenomenon at the origin of the anomalous brightness increase should be
carefully evaluated when dealing with pulsating star models.
\end{enumerate}

It is of paramount 
importance to obtain very accurate light curves 
to give more confidence to these results. However, 
it should be noted that the  $\phi_{31}$ and $\phi_{41}$ values (Tab.~1)
supply a good confirmation of the reliability of the least--squares fits:
indeed, their mean values (1.20 rad and 5.18 rad, respectively) are
in excellent agreement with the expected ones on the basis of the results
on the galactic variables (see Fig.~2 in Antonello et al.
1986, upper and middle panels).

\begin{acknowledgements} The author wishes to thank 
J. Vialle for checking the English form of the manuscript. 
\end{acknowledgements}

\end{document}